\documentclass[10pt]{article}
\usepackage{ametsoc2col}
\newcommand{\myabstract}{Since the beginning of satellite observations, the Arctic sea ice extent has shown a downward trend. The decline has been weaker in the March maximum than in the September minimum and masked by inter-annual fluctuations. One of the less understood aspects of the sea ice response is the persistence times for fluctuations, which could indicate the dominant physical processes behind the sea ice decline. To determine the fluctuation persistence times, however, it is necessary to first filter out the dominant effect of the seasonal cycle. In the current study, we thus develop a statistical model, which accurately decomposes the ice area changes into: (1) a variable seasonal cycle component with a constant shape and (2) a residual (short term) fluctuation. 

We find the persistence time of fluctuations to be only about three weeks, independently from season, which is substantially shorter than previously reported. Such short time scale points to the dominance of atmospheric forcing. The shape of the seasonal cycle is surprisingly constant for the whole observational record despite the rapid decline. This is in agreement with the suggestion that the asymmetry of the seasonal cycle is an effect of Arctic land-sea geography, which has not changed with climate change.

The analysis suggest a jump in the annual sea ice area amplitude occurring in 2007,  from which it has not yet recovered, possibly revealing a permanent amplitude shift. In physical sense, this could imply a shift towards the younger, thinner and more susceptible ice cover commencing after the immense 2007 multi-year ice loss.}
\begin{document}
%
%%%%%%%%%%%%%%%%%%%%%%%%%%%%%%%%%%%%%%%%%%%%%%%%%%%%%%%%%%%%%%%%%%%%%
% TITLE
%
% Enter your TITLE here
%%%%%%%%%%%%%%%%%%%%%%%%%%%%%%%%%%%%%%%%%%%%%%%%%%%%%%%%%%%%%%%%%%%%%
\title{\textbf{\large{Fluctuations and seasonality in the Arctic sea ice area: A sudden regime shift in 2007?}}}
%
% Author names, with corresponding author information. 
% [Update and move the \thanks{...} block as appropriate.]
%
\author{\textsc{Peter D. Ditlevsen}
				\thanks{\textit{Corresponding author address:} 
				Peter D. Ditlevsen, Centre for Ice and Climate, Niels Bohr Institute, 
                               University of Copenhagen, Juliane Maries Vej 30,
				2100 Copenhagen, Denmark. 
				\newline{E-mail: Peter Ditlevsen <pditlev@nbi.ku.dk>}} \\
\textit{\footnotesize{Centre for Ice and Climate, Niels Bohr Institute, University of Copenhagen}}\\
%\and 
\centerline{\textsc{and Ivana Cvijanovic}}\\% Add additional authors, different insitution
\centerline{\textit{\footnotesize{Centre for Ice and Climate, Niels Bohr Institute, University of Copenhagen}}}\\
\centerline{\textit{\footnotesize{and Department of Global Ecology, Carnegie Institution for Science, Stanford}}}\\
}
%
% Formatting done here...Authors should skip over this.  See above for abstract.
\ifthenelse{\boolean{dc}}
{
\twocolumn[
\begin{@twocolumnfalse}
\amstitle

% Start Abstract (Enter your Abstract above.  Do not enter any text here)
\begin{center}
\begin{minipage}{13.0cm}
\begin{abstract}
	\myabstract
	\newline
	\begin{center}
		\rule{38mm}{0.2mm}
	\end{center}
\end{abstract}
\end{minipage}
\end{center}
\end{@twocolumnfalse}
]
}
{
\amstitle
\begin{abstract}
\myabstract
\end{abstract}
\newpage
}
%%%%%%%%%%%%%%%%%%%%%%%%%%%%%%%%%%%%%%%%%%%%%%%%%%%%%%%%%%%%%%%%%%%%%
% MAIN BODY OF PAPER
%%%%%%%%%%%%%%%%%%%%%%%%%%%%%%%%%%%%%%%%%%%%%%%%%%%%%%%%%%%%%%%%%%%%%
\section{Introduction}
Current observations~\citep{cavalieri:2012} show more dramatic decline in the ice area  than predicted by most climate model simulations, despite the major improvements in CMIP5 generation of models ~\citep{stroeve:2012a, wang:2012}. Most dramatic is the decrease in the September minimum sea ice area~\citep{stroeve:2012,cavalieri:2012} with the three lowest ever recorded ice extents occurring in September of 2012, 2007 and 2011, respectively ~\citep{NSIDC:2012}. Ice free summers in Arctic are expected within a few decades ~\citep{stroeve:2007, wang:2009, wang:2012}. 

Locally, expected future decline in the Arctic sea ice area will eminently cause huge ecological and socio-economical disturbances such as ecosystem and biodiversity losses, shifts in traditional lifestyles and culture and the pollution of pristine areas, while providing the opportunities for increased trade, shipping and exploitation of natural resources ~\citep{ACIA:2004, AMAP:2011}. However, remote effects are also anticipated. A number of modeling and observational studies suggest a link between the Arctic sea ice extent and the midlatitude circulation and extreme weather events ~\citep{Petoukhov:2010,Deser:2010, Francis:2012,Liu:2012,Bluthgen:2012} while paleo-studies imply possible remote effects in terms of tropical precipitation shifts ~\citep{ChiangBitz:2005,Brocolli:2006, Cvijanovic:2013}.

The sea ice extent is influenced by a number of atmospheric and oceanic processes with some of the key factors being surface air temperature and radiative flux changes, ocean state, interactions with cloud cover as well as the ocean current and atmospheric circulation changes that facilitate the sea ice export out of the Arctic ~\citep{dickson:2000, serreze:2007,kay:2009, proshutinsky:2009, ogi:2012}.  Atmospheric patterns such as North Atlantic Oscillation (NAO) and the Northern Annular Mode (NAM) are believed to be closely linked to sea ice changes~\citep{rigor:2004, rigor&wallace:2004, Rothrock&Zhang:2005}. The large natural variability and the limitations in understanding the effect of climate change on patterns like the North Atlantic Oscillation (NAO) and the Northern Annular Mode (NAM)~\citep{dickson:2000,rigor:2002} make projections difficult.

An important indicator of physical processes determining the sea ice area is the correlation time for fluctuations ~\citep{blanchard-wrigglesworth:2011}. However, for a signal with a strong deterministic variation, in this case the seasonal cycle, the true statistical correlation can be masked. This can be illustrated on the example of sinusoidal seasonal signal (with added noise), where the autocorrelation will simply reproduce the sinusoidal shape, giving an artificial positive correlation with a lag of 1 year that is unrelated to the residual statistical fluctuations. In order to investigate this part of the signal, which in the following will be denoted the short term fluctuation, the seasonal cycle must be removed. This is not a trivial task, since the seasonal cycle itself is subject to changes. Thus, in order to better understand how the Arctic sea ice responds to climate change, observed variations are analyzed using a new decomposition of the seasonality in the total ice area. 

It is found that the shape of the seasonal cycle is stable, even though the annual mean and the amplitude change substantially. This is in agreement with the resent suggestion that the geography of the Arctic basin, which is constant, mutes the winter sea ice area ~\citep{eisenman:2010}. Temporal analysis of the seasonal amplitude cycle shows a jump to higher amplitude in 2007 that has not been followed by a recovery, indicating a likely shift in the amplitude of the seasonal cycle.

The paper is organized as follows: Satellite data used to obtain the sea ice area is described in Section 2, followed by the description of the statistical method used for the decomposition of the annual cycle (Section 3). Derived time scale for short term fluctuations as well as the change in the annual mean sea ice area and amplitude are discussed in Section 4.  Summary and considerations on possible physical interpretations  are given in Section 5. 

\section{The data}
The sea ice area is derived from NASAs Satellite based Scanning Multichannel Microwave Radiometer~\citep{cavalieri:1996} for the period 1979-present: 1979-1987, SMMR from the NASA Nimbus-7 satellite, 1987-present SSM/I on U.S. Defense Meteorological Satellite Program (DMSP) platforms, with the most recent data (from 2008) using the next-generation SSMIS sensor, replacing SSM/I. For further information see National Snow and Ice Data Center's webpage: http://nsidc.org/data/. The data record analyzed is
obtained from the University of Illinois' cryosphere project webpage ~\citep{thecryospheretoday}. 
The observed record (figure 2, top panel), covers the Arctic Ocean and
surrounding waters in the period 1979-present with daily resolution (every second day prior to 1987). In this analysis the ice area, defined as the area weighted by concentration is used. The ice extent, defined as the area covered by 15\% or more ice is another reported measure. The ice area
is arguably a more relevant climatic variable, while the ice extent data is probably of better quality. 
Thus the following analysis has been repeated using the ice extent (see
Supplement: www.gfy.ku.dk/$\sim$pditlev/seaice-suppl.pdf) with results almost identical to the ones presented for the ice area.

%\subsection{First secondary heading}

%\subsection{Second secondary heading}

%\subsubsection{First tertiary heading}

%\subsubsection{Second tertiary heading}

%\paragraph{First quaternary heading}

%\paragraph{Second quaternary heading}

%\section{Method for Decomposition of the Annual Cycle}

Seasonal variations of the Arctic sea ice area since the start of the satellite measurements (1979) are shown in figure 1  (a). A year round
decline is seen after 2007 with a remarkable decrease in annual mean (red and green curves). 

\begin{figure}[H]
  \noindent\includegraphics[width=19pc,angle=0]{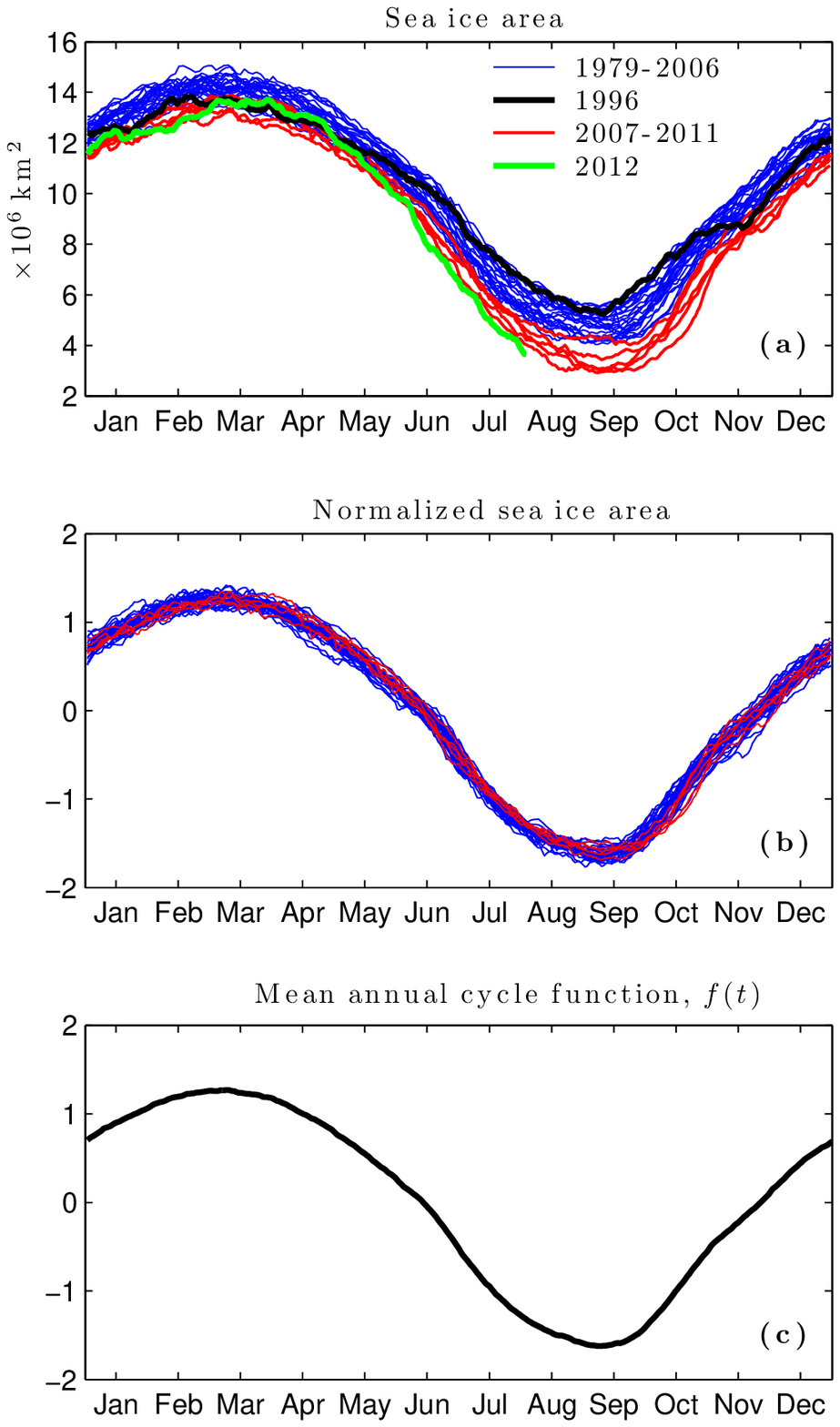}\\
  \caption{The seasonal variation in the Arctic sea ice area. (a) shows all the years obtained from the satellite measurements. A year round decline in is seen after 2007. The year 1996 showed a notably small amplitude in the seasonal cycle. (b) Same data as in (a) but normalized by subtracting annual mean and dividing by the seasonal amplitude (2012 excluded). There is a striking collapse of all years despite the pronounced climate change after 2007. (c) The mean annual cycle function is obtained as the day-by-day average of the 33 normalized curves in (b). }\label{f1}
\end{figure}

\begin{figure}[H] 
  \noindent\includegraphics[width=19pc,angle=0]{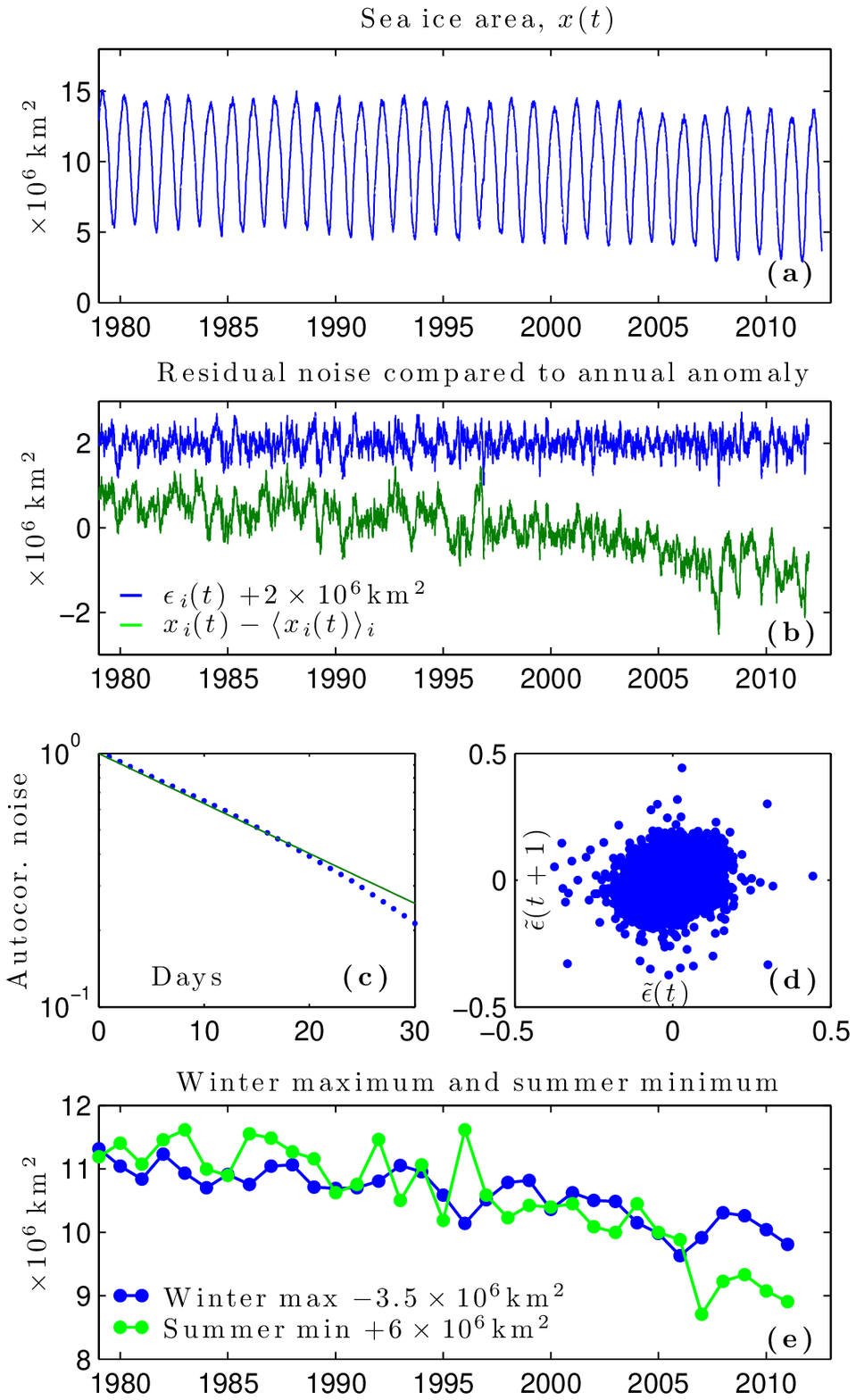}\\
  \caption{(a) shows the Arctic sea ice area measured from  satellites since 1979. (b) shows the short term fluctuation $\epsilon_i(t)$ in blue (shifted upwards for visibility) and the anomaly $x_i(t)-\langle x_i(t)\rangle_i$ in green. The anomaly is not stationary, since it is a mixed signal of both changes in mean and amplitude. (c) shows the autocorrelation of the short term fluctuation, which is almost perfectly exponential. The green line is the curve $\exp(-t/\tau)$, with $\tau=22$ days. (d) The scatter plot of the 
compensated signal (see text for explanation) shows that there is no structure, beside the simple exponential autocorrelation in the short term fluctuation. (e) shows the winter maximum and summer minimum (observe shifted axis for comparison). These are obtained from the mean and amplitude as $min_i=m_i+A_i min(f(t))$ and $max_i=m_i+A_i max(f(t))$ which occur on September 8. and March 9., respectively. }\label{f2}
\end{figure} 

In order to quantify the change in the seasonal cycle of the sea ice area, denoted $x_i(t)$, we decompose it as
\begin{equation}x_i(t)=m_i+A_i f(t)+\epsilon_i(t),\end{equation} 

where $i\in(1979,2011)$ denotes year, and $t\in (1, 365)$ denotes day of year. The mean $m_i$ and amplitude $A_i$ are constant within
a given year $i$. The function $f(t)$ is defined to have zero mean and unit variance. It represents the (constant) seasonal cycle. The residual fluctuation $\epsilon_i(t)$ is assumed in a statistical sense to be a simple stochastic noise. We will refer to $\epsilon_i(t)$ as the short term fluctuation.

It is not given a priory that this is an adequate description of the Arctic sea ice area. However, by plotting the normalized ice area $(x_i(t)-m_i)/A_i=f(t)-\epsilon_i(t)/A_i$ we see an almost perfect collapse of the data (figure 1(b)). Details on how $m_i$ and $A_i$ are calculated are given in the appendix. 

The function $f(t)$ can be accurately estimated as the the mean of the normalized ice area for all years (figure 1(c)).  As the difference between the two lower panels in figure 1 is small, the short term fluctuation $\epsilon_i(t)$ is small. This short term fluctuation is plotted in blue in figure 2(b), (note the scale in comparison to the scale in figure 2(a)).

Two findings are surprising regarding the derived fluctuation $\epsilon(t)$: Firstly, despite the dramatic reduction in Arctic sea ice through the record, there are no trends. Secondly, the short term fluctuation is in a statistical sense indistinguishable from a red noise with a correlation time of $\tau=22$ days. This is seen from the autocorrelation being exponential (figure 2(c)) and the compensated signal

 $\tilde{\epsilon}(t)=\epsilon(t)-\exp(-1/\tau)\epsilon(t+1)$ is perfectly uncorrelated and structureless (figure 2(d)). 

In order to additionally confirm that there is no seasonal dependence, the short term fluctuation it is plotted as a function of time of the year for the 33 year record in figure 3(a). The color coding is as in figure 1, showing no difference between the recent period (2007-2011 in red) and the previous period (1979-2006 in blue).

\begin{figure}[H] 
  \noindent\includegraphics[width=19pc,angle=0]{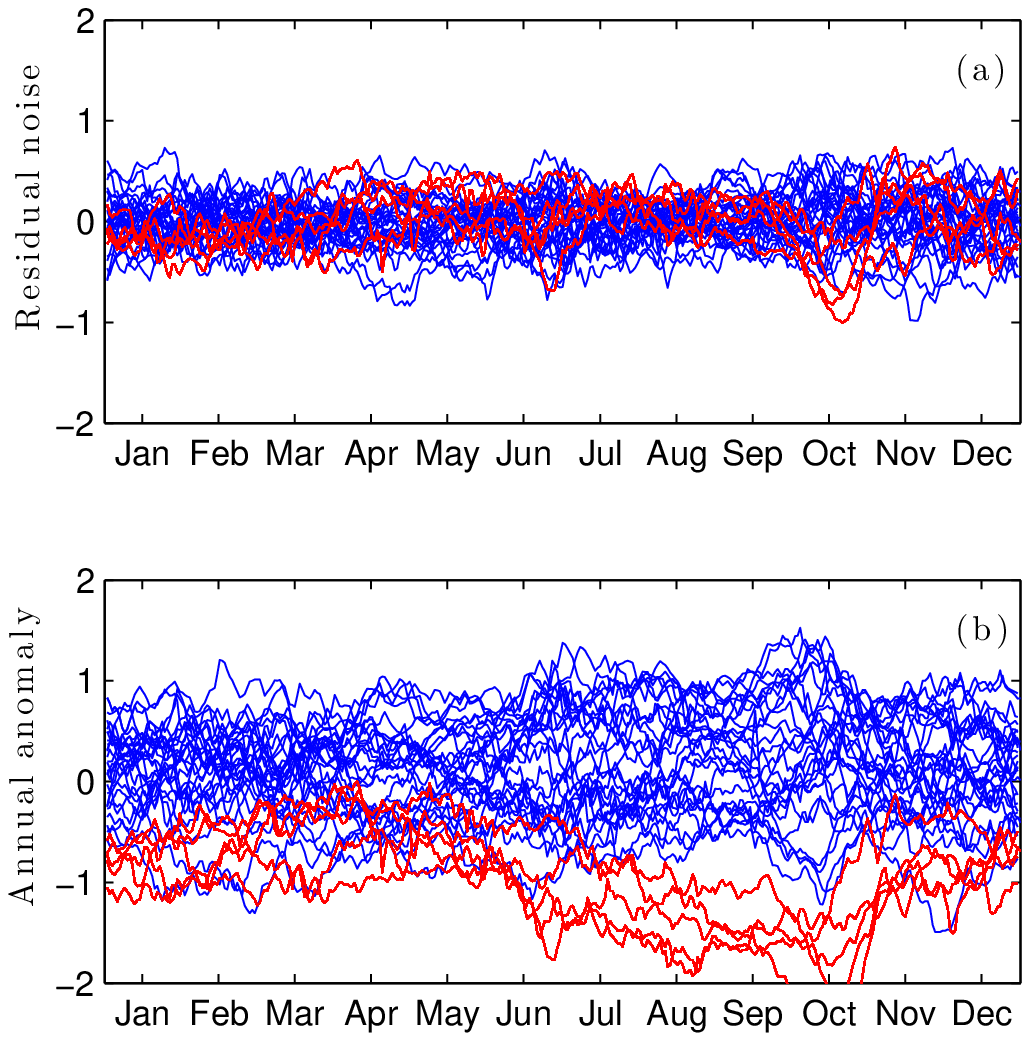}\\
  \caption{The curves in figure 2(b) plotted as function of time of the year. The blue curves are 1979-2006, while the red curves are 2007-2011, as in figure 1. (a) the short term fluctuation for the 33 years shows no seasonality and no statistical difference through the record. This verifies that the seasonality is effectively captured in the first terms in equation (1). (b) the annual anomaly, which shows a difference between the two periods, reflecting the fact that the mean and amplitude of the annual cycle change with time. Especially in the recent period 2007-2011, a seasonal cycle is retained in the anomaly. This will result in an artificial seasonal time scale auto-correlation in the anomaly. }\label{f3}
\end{figure}

Furthermore, following \citep{blanchard-wrigglesworth:2011}, the correlation times calculated from the first day of each month (not shown), also reveals no seasonality in the 
persistence time for the sea ice.

The advantage in considering the short term fluctuation compared to the annual anomaly can be understood when comparing figure 3 (a) and figure 3 (b), where the latter shows the annual anomaly (green curve in figure 2(b)). A clear shift between the early and the late part of the record is seen in figure 3 (b) with a pronounced seasonal cycle retained in the late part of the record. This is a consequence of the mean and amplitude of the annual cycle changing with time. As a further consequence the correlation time in the anomaly record will be on the order of a season.  

The finding of a 22 days correlation time for fluctuations is in contrast to previous findings ~\citep{blanchard-wrigglesworth:2011} of persistence time of several months varying with season. 
Their result is likely a consequence of the persistence in the annual cycle, that is largest at the extremes and smallest at spring and fall, when melting and refreezing results in fast changes in the ice cover. The influence of the annual cycle on the autocorrelation of the sea ice is also noted in \citep{agarwal:2012}.

The correlation time is consistent with the time scale of variations in sea ice area governed by the natural atmospheric variability, while it is short in comparison to the typical persistence times for SST anomalies.

The annual cycle function $f(t)$ is almost identical to the usual annual mean $\langle x_i(t)\rangle_i$ normalized to zero mean and unit variance. Here $\langle x_i(t)\rangle_i$ denotes average for day $t$ in the year over all the years. But the short term fluctuation $\epsilon_i(t)$ is very different from the usual anomaly $x_i(t)-\langle x_i(t)\rangle_i$~\citep{livina:2012,ditlevsen:2012}, which is plotted in green in figure 2(b). The anomaly cannot be considered as a stationary independent noise.      

The sea ice area at any particular day of the year is easily obtained from the decomposition, thus we have the summer minimum $min_i=m_i-1.62 A_i$, where $min(f(t))=f(\mbox{September 8})=-1.62$ and the winter maximum $max_i=m_i+1.27 A_i$, $max(f(t))=f(\mbox{March 9})=1.27$. These are shown in figure 2(e), both with downward trends, most pronounced in the summer minimum. Note that the natural variability unrelated to climate change, represented by $\epsilon_i(t)$ is filtered out of the estimates for $min_i$ and $max_i$.

\begin{figure}[H] 
  \noindent\includegraphics[width=19pc,angle=0]{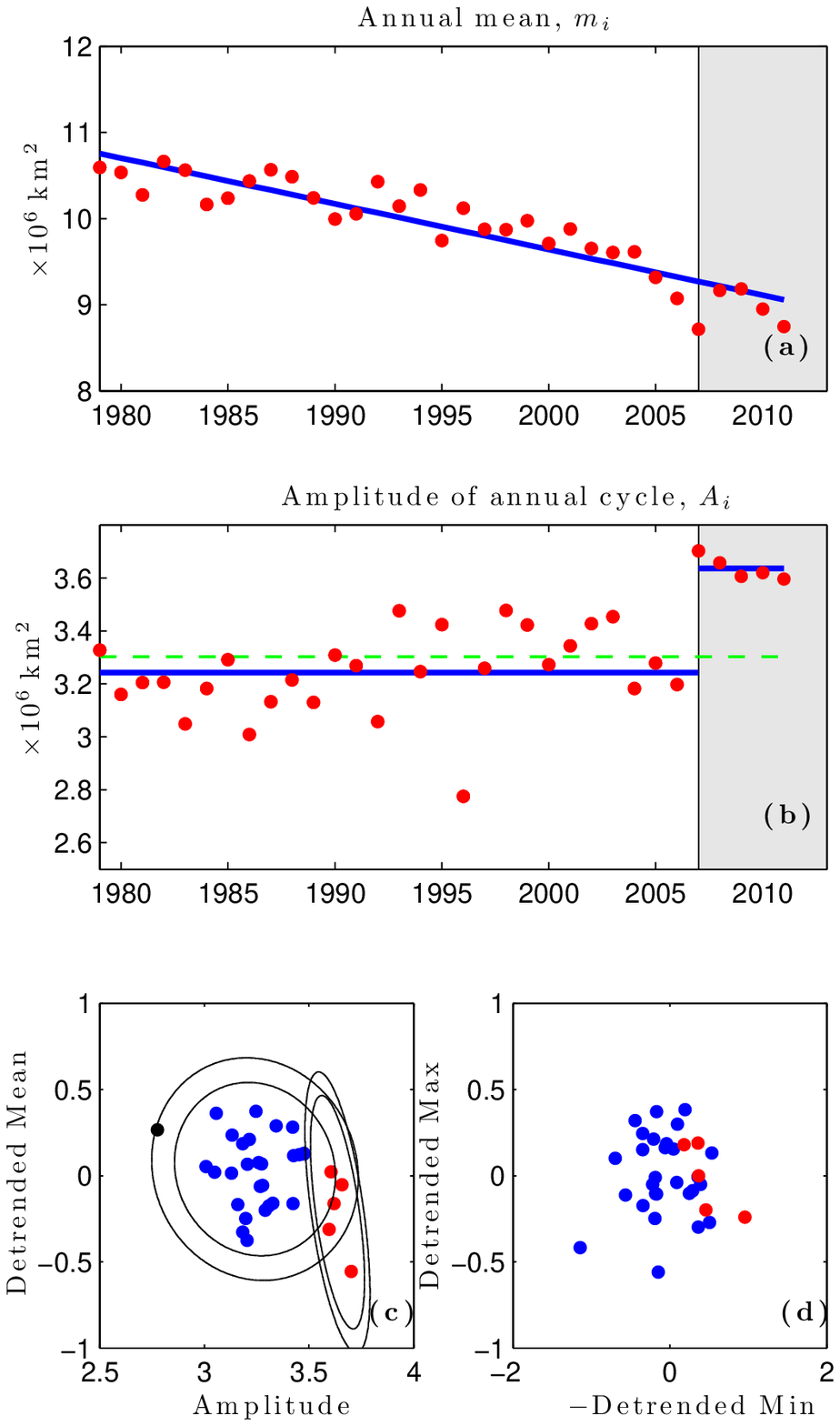}\\
  \caption{The change in statistics in 2007. (a) shows the annual mean $m_i$, with a negative linear trend through the whole record. No statistically significant change in the trend is observed. (b) The change in the amplitude in 2007 is much more significant. The last five points are much higher than the mean for the period (green dashed line) indicating a change in mean (blue lines). (c) Scatter plot of the detrended mean $\tilde{m}_i$ vs. $A_i$ where the color coding is the same as in figure 1(a). This indicates that 2007 is a change point to a new statistical state. Note that $A_i$ and $\tilde{m}_i$ are independent within the two populations, prior to 2007 (blue and black points) and after 2007 (red points).The ellipses show the 90\% and 98\% probability contours for the two maximum likelihood bivariate normal distributions. (d) Shows the scatter of the detrended winter maxima vs. detrended summer minima, these are independent and the period 2007-2011 does not show a significantly different joined distribution in comparison to the earlier part of the record.}\label{f4}

\end{figure}

The annual mean $m_i$ (figure 4(a)), shows, as has been reported before ~\citep{stroeve:2012,lindsay:2009,comiso:2008,maslanik:2007}, a downward trend through the full record, while the amplitude of the seasonal cycle $A_i$ (figure(4(b)) has a distinct minimum for 1996 and a sudden positive jump in 2007, not followed by a recovery, thus possibly indicating a new state. This could be a consequence of the shift towards the one-year ice~\citep{maslanik:2007} and will be further discussed in Section 4. 

%\subsection*{First unnumbered secondary heading}

%\subsubsection*{First unnumbered tertiary heading}

%\paragraph*{First unnumbered quaternary heading}

\section{Results: The annual mean and the amplitude}

The linear trend in the mean ice area $m_i$ is shown in figure 4(a). No significant changes in the linear trend can be detected, though an increased downward trend with low statistical significance has been suggested~\citep{comiso:2008}. 
There is no significant trend in the amplitude of the seasonal cycle either over the period 1979-2006 (figure 4(b)). However, in 2007 there is a jump to a higher level,  which has not been followed by a recovery up to this time. The blue lines in figure 4(b) show the means for the periods 1979-2006 and 2008-2011, while the green dashed line shows the mean for the full record. Denoting the detrended mean by $\tilde{m}_i$, the joined distribution of $(A_i, \tilde{m}_i)$ is shown in the scatter plot in figure 4(c), where the color coding for the points is the same as in figure 1(a). Here a change in joined statistics is observed in 2007.  
The ellipses show the 90\% and 98\% contour lines for the maximum likelihood
bivariate normal distributions for the two populations 1979-2006 (blue points, and the 1996 outlier (in black)) and 2007-2011 (red points). Obviously, the estimate
of the distribution after 2007 is uncertain, since it is based on only five points. The slight tilt in the ellipses around the last five (red) points in figure 4(c) is not significant, thus
there is no significant correlation between the detrended mean and the amplitude. 
The change in amplitude in 2007 is not just a consequence of the summer minimum decreasing more than the winter maximum. Subtracting the trends from the winter maxima and summer minima, the scatter of detrended maxima vs. (minus) the detrended minima is shown in figure 4(d). It is 
clearly seen that the (detrended) yearly minima and maxima are uncorrelated, and that the last part of the record does not show a significantly different distribution. The same
is the case for the summer minimum and the winter maximum of the following year (not shown). Thus the change point in 2007 showing up in the amplitude is masked by the overall trend in summer minimum and winter maximum. 

\section{Discussion and Summary}
In summary, the decomposition (1) of the seasonally varying Arctic sea ice area gives a much clearer insight into the variations than the traditional decomposition into the mean annual cycle and the anomaly, most strongly expressed in the difference between the short term fluctuation and the anomaly in figure 2(b). The correlation time for the fluctuation, that is independent from the annual cycle, is 22 days which points to atmospheric variability as driver, while the ocean state changes and other forcing are reflected in the mean and amplitude of the annual cycle.

The mean $m_i$ shows a steady decline, while the amplitude $A_i$ shows a sudden jump in 2007 as shown in figure 4 (a) and (b). The amplitude and the detrended noise $(A_i, \tilde{m}_i)$ is independent from year to year following a bivariate normal distribution, with a sudden change in distribution in 2007.  

The annual cycle is surprisingly regular, where only the annual amplitude and mean change from year to year. This is remarkable for two unrelated reasons: firstly, the variations in the total ice area includes regions with large variations and a negative trend in the mean.

Secondly, different changing factors influencing the ice growth and retreat have distinct seasonalities~\citep{serreze:2007,lu:2009}. The regularity of the shape of the annual cycle could be in support of the resent suggestion that the geography of the Arctic basin, which is constant, determines the shape by muting the 
winter sea ice area ~\citep{eisenman:2010}. The decomposition of the sea ice area (1), and the constancy of the normalized annual cycle, indicate that all these factors add up such that the annual amplitude is the strongest indicator of climate change.

As there was no recovery in the  amplitude of annual sea ice area after the 2007 jump, we refer to it as a shift into a new state with the increased amplitude of the seasonal cycle.
According to ~\citep{maslanik:2007}, summer of 2007 suffered extreme Arctic ice loss with the ice extent 42 percent
smaller compared to its value in the mid-eighties. The 2007 Artic sea ice loss was unprecedented due to the fact that it was not the increased transport out of the Arctic that was responsible for this record low (as it was the case for the record lows prior to 2007). It was the actual failure of sea ice to survive the Arctic summer making the vast areas north of Alaska and eastern Siberia ice free ~\citep{Stroeveetal:2007}. The atmospheric conditions (persistent anticyclone over the Arctic Ocean during the summer of 2007) and previous years of transition towards the thinner younger ice have majorly contributed to this record low ~\citep{Stroeveetal:2007, maslanik:2007}. The amplitude shift found in this study, thus very likely represents the exact signature of a transition towards the younger, thinner ice that is more susceptible to global warming.

Finally, the described decomposition used in this study could also be used for testing the models in reproducing the described short term fluctuation characteristics as well as the shifts in the amplitude of the seasonal cycle. 
Available daily data briefly tested in this study (output from the Community Climate System Model, version 3 (CCSM3) Large Ensemble Experiment) were not able to reproduce the shifts reported in this study. More comprehensive analysis will be reported elsewhere.

%\begin{acknowledgment} 
%Start acknowledgments here.
%\end{acknowledgment}

% Use appendix}[A], {appendix}[B], etc. etc. in place of appendix if you have multiple appendixes.
\ifthenelse{\boolean{dc}}
{}
{\clearpage}
\begin{appendix}
\section*{\begin{center} Decomposing the sea ice area $x(t)$\end{center}}

By averaging equation (1) over the year we estimate $m_i\approx \langle x_i(t)\rangle_t\pm \sigma/\sqrt{\tilde{n}}$, where $\sigma^2$ is the variance of the
short term fluctuation 
$\epsilon(t)$ and $\tilde{n}=365/22$ is the effective number of independent points within a year. A posterior the relative intensity of 
the short term fluctuation
is calculated to $\sigma/m=0.02$, thus the uncertainty is negligible. Likewise we obtain the amplitude from 
$\langle x_i(t)^2\rangle_t=m_i^2+A_i^2\langle f(t)^2\rangle_t +\sigma^2\Rightarrow A_i=\sqrt{\langle x_i(t)^2\rangle - m_i^2}$, where we have safely neglected the $\sigma^2$ term, which gives a relative error of less than 0.004. The mean cycle function is obtained as $f(t)=\langle (x_i(t)-m_i)/A_i\rangle_i$, where $\langle . \rangle_i$ denotes
averaging over the 33 years. The uncertainty is of the order $\sigma/(\langle A_i\rangle_i\sqrt{33})=0.004$, thus also negligible. As a consistency check of equation (1), we obtain $\langle f(t)\rangle_t/\sqrt{\langle f(t)^2\rangle_t}=0.0015\approx 0$. The effect on the analysis of changing the beginning date for the year is very small (not shown). This has been checked by repeating the full analysis beginning the year in April 1., July 1., October 1. and at winter maximum and summer minimum.    
\end{appendix}

% Create a bibliography directory and place your .bib file there.
% -REMOVE ALL DIRECTORY PATHS TO REFERENCE FILES BEFORE SUBMITTING TO THE AMS FOR PEER REVIEW
\ifthenelse{\boolean{dc}}
{}
{\clearpage}

%\bibliographystyle{ametsoc}
%\bibliography{references}

%%%%%%%%%%%%%%%%%%%%%%%%%%%%%%%%%%%%%%%%%%%%%%%%%%%%%%%%%%%%%%%%%%%%%
% FIGURES-REMOVE ALL DIRECTORY PATHS TO FIGURE FILES BEFORE SUBMITTING TO THE AMS FOR PEER REVIEW
%%%%%%%%%%%%%%%%%%%%%%%%%%%%%%%%%%%%%%%%%%%%%%%%%%%%%%%%%%%%%%%%%%%%%

\end{document}